\documentstyle[aps,preprint,prl]{revtex}
\begin{document}
\draft
\title{Localization and band-gap pinning in semiconductor
superlattices with layer-thickness fluctuations}

\author{Kurt A. M\"ader \cite{cecam} and Alex Zunger}
\address{National Renewable Energy Laboratory, Golden, CO 80401}
\maketitle
%
\begin{abstract}
We consider (AlAs)$_{n}$/(GaAs)$_{n}$ superlattices with random
thickness fluctuations $\Delta n$ around the nominal period $n$. Using
three-dimensional pseudopotential plane-wave band theory, we show that
(i) {\em any} amount $\Delta n/n$ of thickness fluctuations leads to
band-edge wavefunction localization,
(ii) for small $\Delta n/n$ the SL band gap is pinned at the gap level
produced by a {\em single\/} layer with ``wrong'' thickness $n+\Delta n$,
(iii) the bound states due to monolayer thickness fluctuations lead to
significant band-gap reductions, e.g., in $n=2,4,6$, and 10 monolayer
SL's the reductions are 166, 67, 29, and 14 meV for
$\langle111\rangle$ SL's, and 133, 64, 36, and 27 meV for
$\langle001\rangle$ SL's,
(iv) $\langle001\rangle$ AlAs/GaAs SL's with monolayer thickness
fluctuations have a direct band gap, while the ideal
$\langle001\rangle$ SL's are indirect for $n<4$.
\end{abstract}
%
%


\narrowtext

The electronic structure and quantum-confinement effects in
semiconductor superlattices are usually modeled by assuming an {\em
ideal} structure, i.e., that the interfaces are atomically abrupt and
that the individual layer thicknesses remain constant throughout the
superlattice \cite{Bylander86,Wei88}.  Actual heterostructures,
however, often deviate from ideality in two ways: (i) {\em lateral\/}
imperfections in the $(x,y)$ plane, such as intermixed
\cite{Ourmazd89}, stepped \cite{Feenstra94}, or islanded
\cite{Bimberg87} interfaces, and (ii) {\em vertical} ($z$ direction),
discrete thickness fluctuations around its nominal value, even though
the interfaces could remain reasonably flat and atomically abrupt in
the $(x,y)$ plane \cite{Fujiwara87}.  It is likely that both types of
interfacial imperfections coexist in many samples \cite{Warwick90}.
In case (i) the translational symmetry is broken {\em in the substrate
plane\/} $(x,y)$, and the concentration profile along the growth
direction is continuous (``graded'' or ``intermixed'' interfaces),
while in case (ii) a discrete, rectangular-shaped concentration
profile exists, but the superlattice translational symmetry is broken
{\em along the growth direction\/} $z$.  Case (i) has been modeled
theoretically by assuming graded (rather than square) potential wells
\cite{Nelson87}, or by considering supercells with atomic swaps across
the interfaces \cite{Laks92}.
In this paper we consider layer-thickness fluctuations [case (ii)],
i.e., $(A)_{n}/(G)_{n}$ superlattices (SL) with nominal layer
thicknesses of $n$ monolayers (ML) of material $A=$ AlAs and $n$ ML of
material $G=$ GaAs, but where each layer is occasionally thinner or
thicker than its intended thickness $n$.  We use a three-dimensional
pseudopotential band-structure description within a highly flexible
plane-wave basis \cite{Mader94}, rather than one-dimensional
effective-mass models \cite{Littleton86}, or one-dimensional
\cite{Dow82} or three-dimensional \cite{EGWang94} tight-binding
models.

Superlattices with layer-thickness fluctuations are described by
supercells containing several hundreds of atoms \cite{Wang94}. The
novel empirical pseudopotentials used here \cite{Mader94} have been
tested extensively for AlAs/GaAs bulk materials, short-period
superlattices, and random alloys. The results \cite{Mader94} compare well with
experiment and with state-of-the-art, self-consistent pseudopotential
calculations, however, without suffering from the band-gap
underestimation of the local-density approximation \cite{Mader94}.
For periods $n\le20$ we consider {\em single monolayer fluctuations},
so the layer thicknesses are in the set
$\{n-1,n,n+1\}$, while for $n>20$ we
consider a fixed fraction $\Delta n / n$ of layer-thickness
fluctuations $\Delta n$.

We find that:
(i) {\em any} amount $\Delta n/n$ of thickness fluctuations leads to
band-edge wavefunction localization,
(ii) for small $\Delta n/n$ the SL band gap is pinned at the gap level
produced by a {\em single\/} layer with ``wrong'' thickness $n'
\not=n$ (a ``chain mutation''),
(iii) the bound states due to monolayer thickness fluctuations lead to
band-gap reductions that monotonically decrease with increasing $n$.
These fluctuation-induced bound states
will photoluminesce at energies below the ``intrinsic'' absorption
edge.
(iv) $\langle001\rangle$ AlAs/GaAs SL's with monolayer thickness
fluctuations have a direct band gap, while the ideal
$\langle001\rangle$ SL's are indirect for $n<4$.

Consider first, by way of reference, the band structure of {\em
ideal\/} AlAs/GaAs SL's with layers oriented along (111) or (001)
(Fig.~\ref{fig:gaps}). In agreement with previous theoretical studies
\cite{Wei88}, we find that
(i) the ideal $\langle111\rangle$ SL has a {\em direct\/} band gap for
all $n$ values, since the conduction-band minimum (CBM) is the
$\Gamma$-folded $\bar\Gamma_{\rm 1c}(\Gamma_{\rm 1c})$ state. This SL
has a ``type-I'' band arrangement with both the highest valence and
the lowest conduction state localized on the GaAs layers.
(ii) The second conduction band at $\bar\Gamma$ is folded from the
zincblende $\Gamma$--$L_{z}$ bands; for small $n$ the pseudodirect
$\bar\Gamma_{\rm 1c}(L_{\rm 1c})$ state mixes strongly with the direct
$\bar\Gamma_{\rm 1c}(\Gamma_{\rm 1c})$ state. The mixing, and thus the level
repulsion, shows odd-even oscillations for small $n$ (reflecting
localization of repelling states on the same or on either sublattice
\cite{Wei88}).
Note that the one-dimensional effective-mass model (dashed
lines \cite{Wang95} in Fig.~\ref{fig:gaps}) completely misses the strong
non-monotonic variations of SL energy levels with layer
thickness.

The situation is very different for $\langle001\rangle$ oriented ideal
SL's.  The prominent properties apparent in Fig.~\ref{fig:gaps}(b)
are:
(i) the $n=1$ SL has an {\em indirect\/}
band gap at the $L$-folded point $\bar R$ \cite{Bylander86,Wei88};
(ii) for $n<4$, the lateral $X_{x,y}$ valleys (folded to $\bar M$) and
the $X_z$ valley (folded to $\bar\Gamma$) are nearly degenerate
\cite{Ge94};
(iii) for $1<n\le8$ the pseudodirect, AlAs-like
$\bar\Gamma_{\rm 1c} (X_z)$ state is below the direct, GaAs-like
$\bar\Gamma_{\rm 1c} (\Gamma_{\rm 1c})$ state, thus the SL is type II;
for $n>8$, however, $\bar\Gamma_{\rm 1c} (\Gamma_{\rm 1c})$ is lower,
so the system is type I (experimentally, the type-II/type-I crossover
is found at $n\approx11$ \cite{Ge94});

We now allow the layer thicknesses in the $n\times n$ SL's to
fluctuate around the ideal value $n$ by $\pm1$ ML\@. The growth
sequence is now defined in terms of a {\em distribution function\/}
$p(n')$, which we assume to be uncorrelated and symmetric around the
nominal thickness $n$. We define the relative frequency $R$ of
monolayer fluctuations by
\begin{equation}
\label{eq:R}
	p(n\pm1) = R\, p(n).
\end{equation}
Because the distribution $p(n')$ is normalized, we can write $p(n) =
1/(1+2R)$ and $p(n\pm1) = R/(1+2R)$.  For the {\em ideal\/} $n\times
n$ superlattice $R=0$, and $p(n') = \delta (n'-n)$, whereas for $R=1$
the layer thicknesses $\{n-1,n,n+1\}$ occur with equal probability $p
= \frac{1}{3}$.  A single chain mutation in a finite superlattice of
length $N$ monolayers corresponds to $R\approx\frac{2n}{N}$, which
will denoted here as the $R\to0$ limit (to be distinguished from $R=0$
with no mutations).  To simulate the lack of periodicity along the
growth direction, we have used supercells of a total length $N$
between 100 and 1000 ML, and repeated these supercells periodically.
A particular growth sequence was created by a random number generator,
specifying $n$, $N$ and $R$.

The band-edge energies of (AlAs)$_{n}$/(GaAs)$_{n}$
SL's with {\em one-monolayer thickness fluctuations\/} about $n$
are plotted in Fig.~\ref{fig:bind} relative to the band edges of the
{\em ideal\/} SL's (the energy zero).
We see that:
(i) thickness fluctuations create both hole ($\Delta\varepsilon_{\rm
h}$) and electron ($\Delta\varepsilon_{\rm e}$) bound states for {\em
any degree\/} of thickness fluctuation ($0<R\le1$),
(ii) the band-gap reductions $\Delta E_{\rm g} =
\Delta\varepsilon_{\rm h} + \Delta\varepsilon_{\rm e}$ decay with $n$,
and have a definite dependence on superlattice direction; they are
166, 67, 29, and 14 meV for $n=$2,4,6, and 10 in the
$\langle111\rangle$ direction, and 133, 64, 36, and 27 meV in the
$\langle001\rangle$ direction, respectively,
(iii) the dilute  limit of a single chain mutation already produces a
{\em finite gap reduction} $\Delta E_{\rm g}(R\to0)$,
(iv) $\Delta E_{\rm g}(R\to0)$ merges with $\Delta E_{\rm g}(R=1)$ at
$n\gtrsim6$, at which point the gap reduction becomes independent
(``band-gap pinning'') of the number of chain mutations.

The appearance of gap levels in SL's with one-monolayer thickness
fluctuations is accompanied by {\em wavefunction localization}. For
example, inspection of the CBM wavefunction of an $n=6$
$\langle111\rangle$ superlattice with random $\pm1$ monolayer
fluctuations [Fig.~\ref{fig:psi}(a)] reveals that it is localized on
$\sim$4 GaAs wells, with minimal amplitude in the AlAs barriers and
maximal amplitude on the two neighboring mutated (7-ML) GaAs wells
(``twin'' fluctuation denoted by bold arrows). The CBM thus resembles
a bound state in a coupled double quantum well. The hole wavefunction at the
valence-band
maximum (VBM) is likewise localized on a number of mutated, 7-ML GaAs
wells [Fig.~\ref{fig:psi}(a)]; in contrast to the CBM, however, the
reason for the multi-well pattern of the VBM wavefunction is that
these states are in fact decoupled, quantum-well confined states,
which are degenerate in energy within the accuracy of our calculation
($\lesssim$0.1 meV).  A typical hole
and electron wavefunction localized on an {\em isolated} (GaAs)$_{7}$
mutation in an otherwise ideal $6\times6$ $\langle111\rangle$ SL are
shown in Fig.~\ref{fig:psi}(b). We see that the hole wavefunction of
an isolated mutation ($R\to0$) resembles that of the concentrated
($R=1$) mutations [Fig.~\ref{fig:psi}(a)], and its binding energy
$\Delta\varepsilon_{\rm h}(R\to0)= 11$ meV equals the value at $R=1$.
At the CBM, the larger penetration of the wavefunction into
neighboring GaAs wells can produce deeper gap states, and
consequently pinning occurs at a larger $n$ ($n_{\rm p}\approx10$)
than for hole states.

Experimentally, the fluctuation-induced localized bound states should
be observable as photoluminescence centers whose energy is below the
absorption edge of the underlying ``ideal'' SL structure.  This
photoluminescence will lack phonon lines, because the optical
transitions are direct in the planar Brillouin zone (the transverse
wave vector $\vec k_{\perp}$ is still a good quantum number), and
because the $k_z$ selection rule is relaxed by vertical disorder.
Indeed, while calculations \cite{Wei88} on {\em ideal\/}
$\langle111\rangle$ (AlAs)$_{n}$/(GaAs)$_{n}$ SL's predicted a direct
band gap with a type-I band arrangement, Cingolani et al.\
\cite{Cingolani90} noted a $\sim$100 meV red shift of the
photoluminescence at 1.80 eV relative to the absorption in
(AlAs)$_{6}$/(GaAs)$_{6}$ $\langle111\rangle$ SL's, interpreting this
as reflecting a type II band arrangement. However, since they noted
that their SL had a $\pm1$ ML period uncertainty, it is possible that
the red shifted photoluminescence originates from
thickness-fluctuation bound states. Our calculated band gap of the
$n=6$ superlattice with $\pm1$ ML thickness fluctuations is 1.78 eV
for $R=1$, and 1.80 eV for $R\to0$, close to their observed
photoluminescence peak position (1.80 eV) \cite{Cingolani90,mix}.


Figure~\ref{fig:bind} shows that the bound states of {\em isolated}
mutations ($R\to0$) merge with those of concentrated layer-thickness
fluctuations ($R\to1$) at some ``pinning period'' $n_{\rm p}$. In what
follows we discuss the (i) $n<n_{\rm p}$ and (ii) $n\ge n_{\rm p}$
regimes.  In the short-period regime ($n<n_{\rm p}$), the band-gap
reduction depends on $R$.  This regime coincides with the region in
Fig.~\ref{fig:gaps} where the band gaps of the {\em ideal\/} SL's have
a complex $n$-dependence, showing strong non-effective-mass behavior.
The $\langle001\rangle$ (AlAs)$_{2}$/(GaAs)$_{2}$ SL with monolayer
fluctuations is in fact identical to the {\em intentionally
disordered\/} SL grown by Sasaki et al.\ \cite{Sasaki89}.  In that
structure, $A_{2}$ and $G_{2}$ layers are randomly replaced by $A_{1},
A_{3}, G_{1}$, and $G_{3}$ layers.  We find the following changes
in the band structure when the layer thicknesses
fluctuate by $\pm1$ ML:
(i) For the $n=2$ $\langle001\rangle$ SL we obtain
$\Delta\varepsilon_{\rm e}(R=1) = $ 22, 81, and 171 meV at $\bar M,
\bar\Gamma$, and $\bar X$ in the planar Brillouin zone.  Since the
level shift $\Delta\varepsilon_{\rm e}$ at $\bar\Gamma$ exceeds the
one at $\bar M$ by $\sim$60 meV, layer-thickness fluctuations
transform the indirect $2\times2$, ideal superlattice into a
direct-gap material \cite{Mader94b}.  The large binding energy
at $\bar X$ originates from the increased level repulsion of the
folded $L_{\rm 1c}$ states when the translational and rotational
symmetry of the ideal $n=2$ superlattice is broken.  This level
repulsion is larger for odd $n$ than for even $n$, and it leads to an
$L$-like, indirect CBM for $n=1$ [Fig.~\ref{fig:gaps}(b)].
(ii) The $n=2$ $\langle001\rangle$ SL is direct even in the $R\to0$
limit of isolated chain mutations.
(iii) Numerous electron and hole states are localized on the same spatial
region along $z$, hence the $\langle001\rangle$ SL's with
thickness fluctuations will exhibit type-I, rather
than type-II characteristics as ideal $n<10$ SL's do.
(iv) While for $n>n_{\rm p}$ only mutated, wider GaAs wells
bind a carrier (see below), in the short-period $\langle001\rangle$
case even an (AlAs)$_{n+1}$ mutation binds an electron. In fact, the
AlAs-like bound electron lies {\em deeper\/} in the gap than the
GaAs-like bound electron [see the two dotted lines near the CBM in
Fig.~\ref{fig:bind}(b)].

We next discuss the SL properties in the regime of
band-gap pinning $n\ge n_{\rm p}$.  At this point the band-gap
reduction is pinned at the value
\begin{equation}
\label{eq:pin}
   \Delta E_{\rm g}(R) = \lim_{R\to0} \Delta E_{\rm g}(R) =
     \Delta\varepsilon_{\rm e} + \Delta\varepsilon_{\rm h},
\end{equation}
where $\Delta\varepsilon_{\rm e}$ ($\Delta\varepsilon_{\rm h}$) is the
electron (hole) binding energy of an {\em isolated\/} ($R\to0$) layer
mutation.  Qualitatively, Eq.~(\ref{eq:pin}) can be understood in
terms of a one-dimensional effective-mass picture, where the SL is
modeled by a Kronig-Penney model, with the usual boundary conditions
of a continuous envelope function $F(z)$ and current $\frac{1}{m^*}
F'(z)$, where the effective mass $m^*$ is different in the well and
barrier material.  Each of the $(n+1)$-ML mutations gives rise to a
bound state below the band edge of the $n\times n$ SL
\cite{Littleton86,Chomette86}.  The gray lines in Fig.~\ref{fig:bind}
indicate that this simple picture is expected to agree {\em
quantitatively\/} with our pseudopotential calculation for
$n\gtrsim10$ (note, however, the large discrepancy with our
three-dimensional calculation at smaller $n$'s).  For very large $n$,
when the quantum wells are completely decoupled, the SL energy
spectrum is simply that of degenerate single quantum wells.  Hence
the extra binding energy of an $(n+\Delta n)$-ML mutation approaches
asymptotically
\begin{equation}
\label{eq:well}
\Delta\varepsilon_{c} =  \varepsilon_0(n) - \varepsilon_0(n+\Delta n) \approx
	\frac{2\Delta n}{n} \, \varepsilon_0(n),
\end{equation}
where $\varepsilon_0(n)$ is the ground-state energy of a carrier with
mass $m^{*}$ in a $n$-ML wide quantum well, which scales like
$\frac{1}{m^{*} n^2}$ for large $n$.  Using a fixed $\Delta n/n =
10$\% we obtain from the first equality of Eq.~(\ref{eq:well})
$\Delta\varepsilon_{\rm e} = 10.0, 2.4$ and 0.7 meV for $n=20, 50$ and 100
in the $\langle111\rangle$ SL
[the last equality of Eq.~(\ref{eq:well}) gives 14.3, 3.0 and
0.8 meV, respectively].  The band-gap reduction for a given $\Delta
n/n$ is obtained by inserting $\Delta\varepsilon_{\rm h}$ and
$\Delta\varepsilon_{\rm e}$ from Eq.~(\ref{eq:well}) in
Eq.~(\ref{eq:pin}).
The degeneracy of the gap level $\Delta\varepsilon_{c}$ is equal to
the number $M_{n+\Delta n}$ of ($n+\Delta n$)-ML well mutations,
which, in the case of $\Delta n =1$ [see Eq.~(\ref{eq:R})], is given
by
\begin{equation}
\label{eq:count}
	M_{n+1} \approx \frac{R}{1+2R}\; \frac{N}{2n}.
\end{equation}
Using Eqs.~(\ref{eq:well}) and (\ref{eq:count}) one can predict the
band-gap reduction and number of localized states in the multiple
quantum-well regime.

In summary, we have shown that discrete layer-thickness fluctuations
in (AlAs)$_{n}$/(GaAs)$_{n}$ superlattices lead to (i) localization of
the wavefunctions near the band edges, (ii) a reduced band gap, which
is pinned at the value corresponding to an isolated layer-thickness
fluctuation for $n>n_{\rm p}$, and (iii) a
crossover to a direct band gap in the case of short-period
$\langle001\rangle$ SL's.

Acknowledgment---We wish to thank L.-W. Wang for discussions and for
performing the envelope-function model calculations of
Figs.~\ref{fig:gaps} and \ref{fig:bind}.  This work was supported by
the Office of Energy Research, Materials Science Division, U.S.\
Department of Energy, under grant No.\ DE-AC36-83CH10093.

%
%

%
%

%
%

\mediumtext
\begin{figure*}
\caption{%
Energy levels of {\em ideal\/} (AlAs)$_{n}$/(GaAs)$_{n}$ superlattices
along (a) $\langle111\rangle$ and (b) $\langle001\rangle$, as a
function of period $n$. The bulk levels in the middle column are
reached asymptotically as $n\to\infty$.  $\Gamma_{\rm 15v}$(AlAs) is
500 meV below the GaAs VBM.  The SL states
are denoted with an overbar, while their parent zincblende states
$\Gamma$, $X$, and $L$ are given parentheses.
The dashed lines are obtained from a
one-band envelope-function model using the effective masses and band
offset from our pseudopotential calculation \protect\cite{Wang95}.
}
\label{fig:gaps}
\end{figure*}

\mediumtext
\begin{figure*}
\caption{%
Gap levels in the presence of one-monolayer thickness
fluctuations in (AlAs)$_{n}$/(GaAs)$_{n}$ superlattices along (a)
$\langle111\rangle$ and (b) $\langle001\rangle$, as a function of period
$n$. Energies are measured with respect to the band extrema of the
{\em ideal\/} $n\times n$ SL (see Fig.~\protect\ref{fig:gaps}).
One-band envelope-function results for a $(n+1)$-ML mutation
embedded in the $n\times n$ SL are indicated by a dashed gray line.
$R=1$ and $R\to0$ denote, respectively, the concentrated and dilute
limit of chain mutations [Eq.~(\protect\ref{eq:R})].
}
\label{fig:bind}
\end{figure*}

\narrowtext
\begin{figure}
\caption{%
Planar averages of wavefunctions squared of the CBM and VBM in
the (AlAs)$_{6}$/(GaAs)$_{6}$ SL along $\langle111\rangle$ with $\pm1$
layer-thickness fluctuations. Hole wavefunctions are plotted in the
negative direction, with a small offset for clarity.  (a) Concentrated
limit ($R=1$), (b) dilute limit ($R\to0$), i.e., a single (GaAs)$_{7}$
mutation embedded in a $6\times 6$ SL host.  The rectangular lines
show the growth sequence of the SL, with GaAs layers represented by
wells, and AlAs layers represented by barriers, respectively.  The
vertical arrows in (a) indicate the 7 ML thick, ``mutated'' wells.  }
\label{fig:psi}
\end{figure}

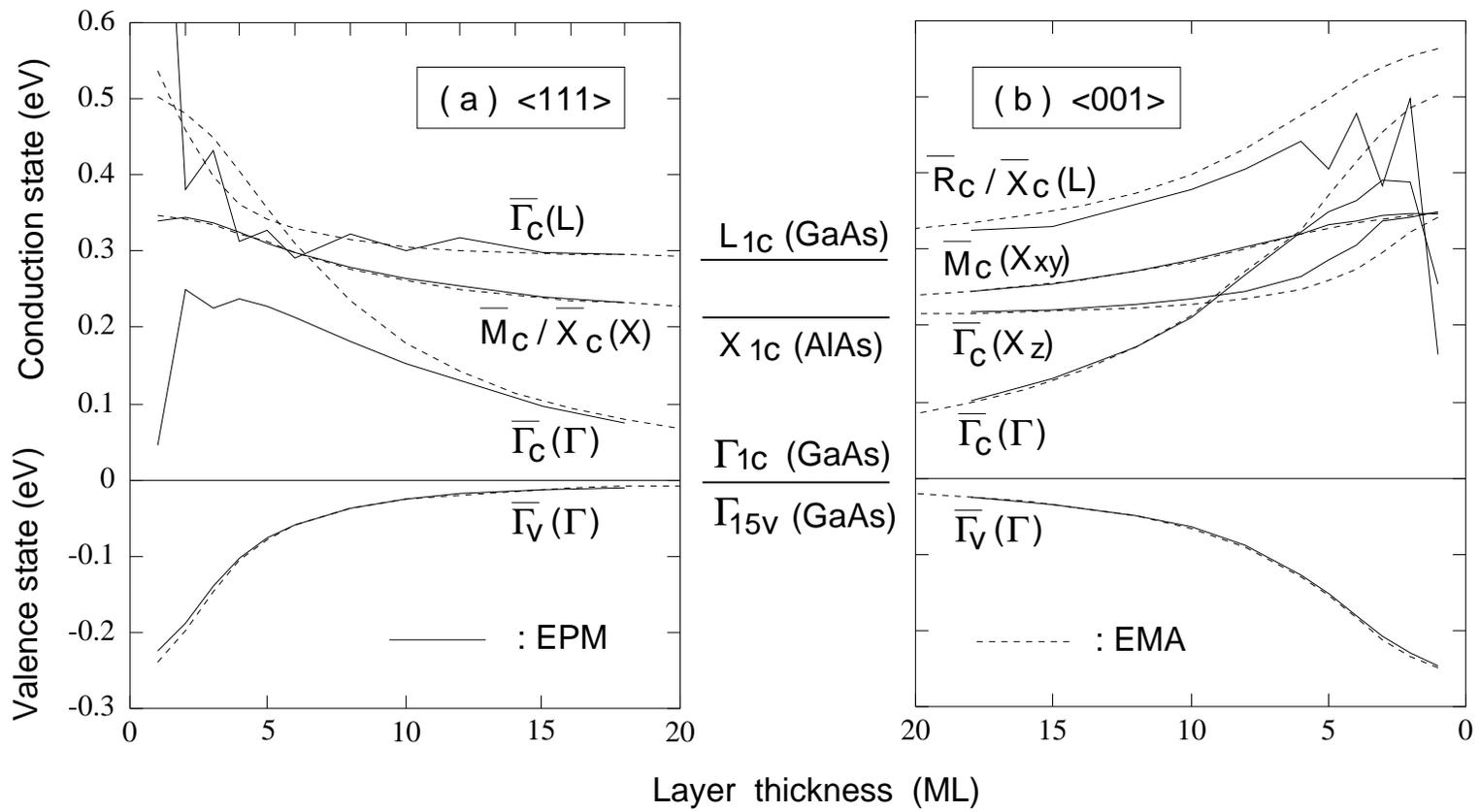
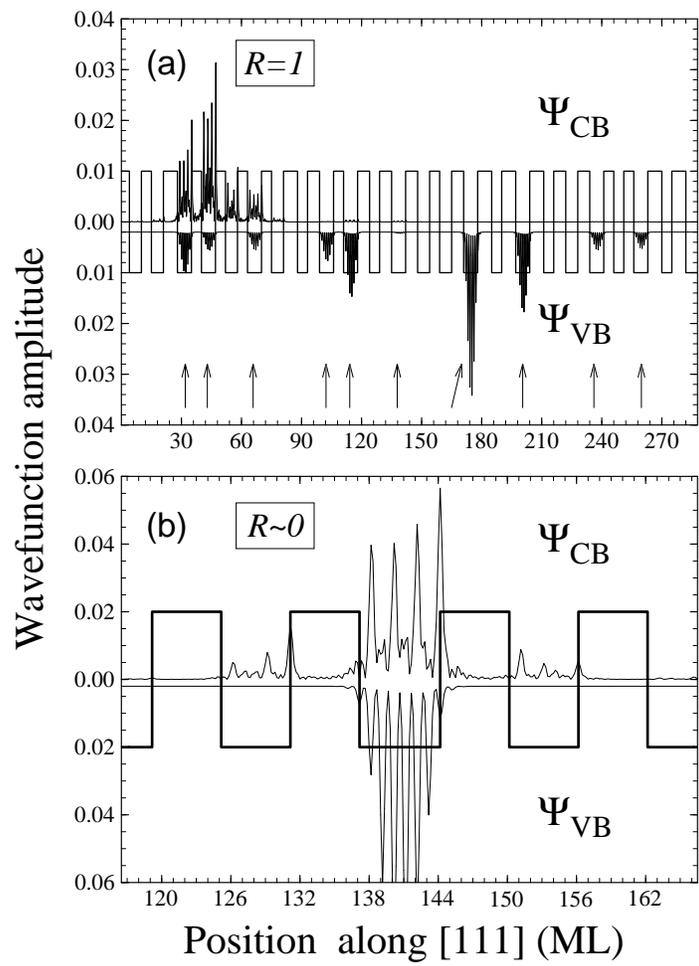
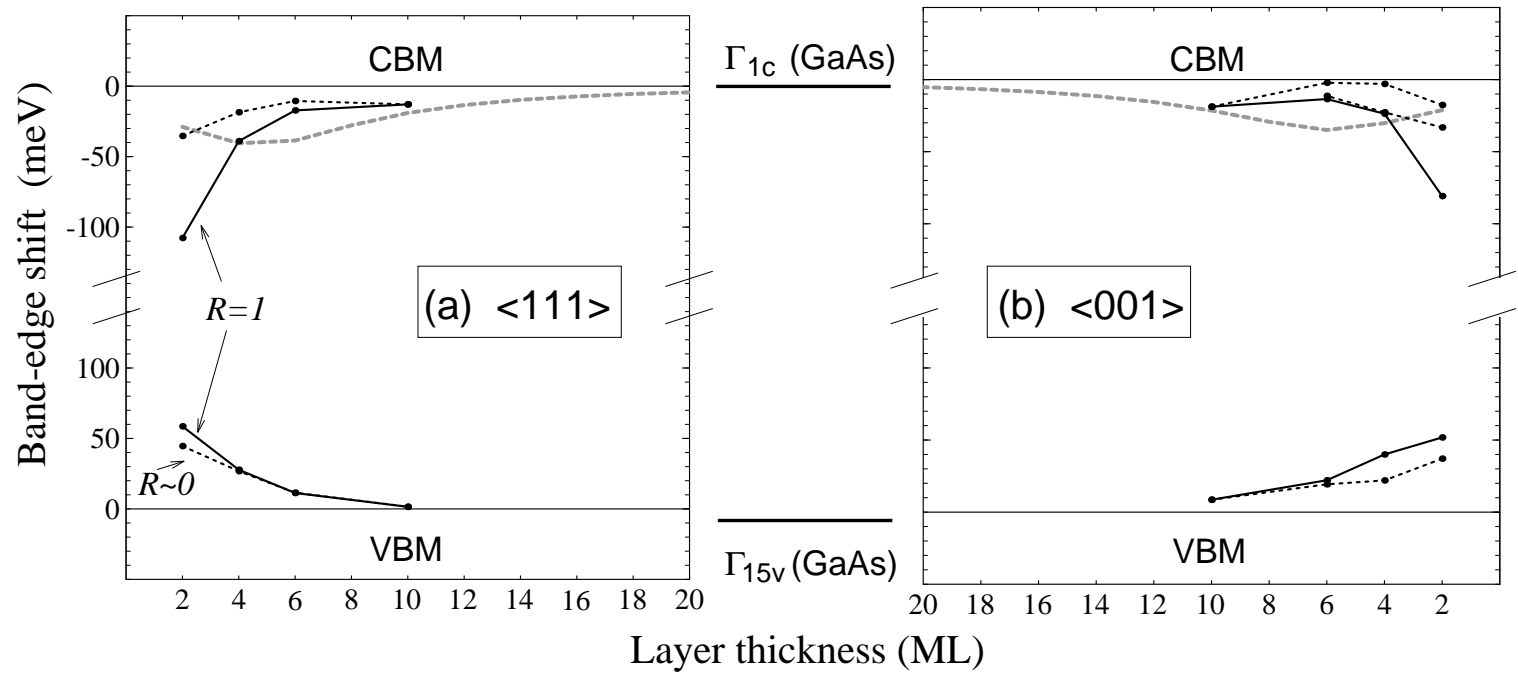
\end{document}